\documentclass[draft]{article}
\usepackage{amssymb}

\newcommand{\Z}{{\mathbf Z}}

\newcommand{\Tr}{\mathrm{Tr}}
\newcommand{\Norm}{\mathrm{N}}

\newcommand{\kl}{\mathrm{ Kl}} % Kloosterman sum
\newcommand{\nmk}{\mathcal{ KT}} % Katz's exponential sum

 \newcommand{\beq}{\begin{equation}}
 \newcommand{\eeq}{\end{equation}}
 \newcommand{\beqa}{\begin{eqnarray}}
 \newcommand{\eeqa}{\end{eqnarray}}

 \def\<{\langle}
 \def\>{\rangle}

 \def\opone{\leavevmode\hbox{\small1\kern-3.8pt\normalsize1}}
 
 \newcommand{\complex}{{\kern .1em {\raise .47ex\hbox {$\scriptscriptstyle
 |$}}\kern -.4em {\rm C}}}
 \newcommand{\real}{{{\rm I} \kern -.19em {\rm R}}}

\begin{document}

\title{Random matrix theory, the exceptional Lie groups, and
$L$-functions.}
\author
{J.P. Keating${}^a$, N. Linden${}^a$ and Z. Rudnick ${}^b$\\
{\protect\small\em ${}^a$School of
Mathematics, University of Bristol, Bristol BS8 1TW, UK.}\\
{\protect\small\em ${}^b$Raymond and Beverly Sackler School of
Mathematical Sciences, }\\{\protect\small\em Tel Aviv University,
Tel Aviv 69978, Israel.} }
\date{31st July 2002}

\maketitle

\abstract {There has recently been interest in relating properties
of matrices drawn at random from the classical compact groups to
statistical characteristics of number-theoretical $L$-functions.
One example is the relationship conjectured to hold between the
value distributions of the characteristic polynomials of such
matrices and value distributions within families of $L$-functions.
These connections are here extended to non-classical groups.  We
focus on an explicit example: the exceptional Lie group $G_2$. The
value distributions for characteristic polynomials associated with
the 7- and 14-dimensional representations of $G_2$, defined with
respect to the uniform invariant (Haar) measure, are calculated
using two of the Macdonald constant term identities.  A one
parameter family of $L$-functions over a finite field is
described whose value distribution in the limit as the size of
the finite field grows is related to that of the characteristic
polynomials associated  with the 7-dimensional representation of
$G_2$.  The random matrix calculations extend to all exceptional
Lie groups.}

 %{\openup 12pt
%\tableofcontents
\newpage

\section{Introduction}

Most work on the connection between random matrix theory and
$L$-functions has concentrated on random matrices chosen from
ensembles related to the {\it classical} compact groups.
Montgomery \cite{Mont}, Rudnick and Sarnak \cite{Rud-Sar}, and
Bogomolny and Keating \cite{Bog-Keat1, Bog-Keat2} calculated the
correlation functions of the zeros of the Riemann zeta-function,
scaled to have unit mean spacing, in the limit as $T$, the extent
of the averaging range up the critical line, tends to infinity.
Their results suggest that these correlation functions coincide
with ones relating to the eigenvalues of unitary matrices in the
limit as the matrix size, $N$, tends to infinity.  In the latter
case the average is defined with respect to the uniform invariant
(Haar) measure on the unitary group $U(N)$; that is, with respect
to the Circular Unitary Ensemble (CUE) of Random Matrix Theory
(RMT). There is extensive numerical evidence in support of this
connection \cite{Odlyzko}, which is expected to extend to the
zeros of any given principal $L$-function. Katz and Sarnak
\cite{KS} conjectured that the distributions of low-lying zeros in
families of $L$-functions are the same as those of the eigenvalues
of matrices from the various classical compact groups (e.g.~the
orthogonal group $O(N)$ and the symplectic group $USp(2N)$, as
well as $U(N)$), the particular group in question being determined
by the symmetry of the family.  This is also supported by
numerical evidence \cite{Rubinstein}.

It was suggested by Keating and Snaith \cite{Keating-Snaith1} that
the value distribution of a given principle $L$-function on its
critical line coincides, in the limit as $T \rightarrow \infty$,
with the value distribution of the characteristic polynomials of
random unitary matrices, defined again by an average with respect
to Haar measure for $U(N)$, in the limit as $N \rightarrow
\infty$.  The random matrix value distribution was calculated in
\cite{Keating-Snaith1} by expressing the group average in terms of
an integral over the matrix eigenvalues, using a formula for the
measure due to Weyl \cite{Weyl}, and then relating the resulting
($N$-dimensional) integral to one evaluated by Selberg.  This idea
was later extended in line with the Katz-Sarnak philosophy to
relate the value distribution within a given family of
$L$-functions at the  centre of the critical strip to the value
distribution of the characteristic polynomials associated with
elements of the appropriate classical compact group in the $N
\rightarrow \infty$ limit \cite{Keating-Snaith2, Conrey-Farmer,
CFKRS}.  Again, the random matrix calculations were performed
using Weyl's integration formula and the Selberg integral.  One
interesting feature of these calculations is that in all cases the
logarithm of the characteristic polynomial, normalized
appropriately (by $\log N$), satisfies a central limit theorem in
the limit $N \rightarrow \infty$. This is in agreement with a
theorem of Selberg which states that the logarithm of the Riemann
zeta function, normalized appropriately (by $\log\log T$), also
satisfies a central limit theorem in the limit $T \rightarrow
\infty$.  For further related developments see \cite{Hughes1,
Hughes2, CKRS}.

It is in this context that we now ask whether there is a
connection between $L$-functions and random matrices from the {\em
non-classical} groups.  A particularly interesting class of these
groups, closely related to the classical groups, is that of the
exceptional Lie groups.  Our purpose in this note is to point out
that a number of the key constructions which serve to provide the
link with random matrix theory in the classical case have
analogues for the exceptional Lie groups.  We illustrate this by
computing the moments and value distribution of the characteristic
polynomials of matrices associated with the 7- and 14-dimensional
representations of one particular exceptional Lie group, $G_2$.
The methods employed are again the appropriate Weyl integration
formula and generalizations of the Selberg integral conjectured by
Macdonald (and known as Macdonald constant term identities)
\cite{Macdonald}, proven for $G_2$ by Zeilberger \cite{Zeilberger}
and Habsieger \cite{Habsieger} and  by Opdam \cite{Opdam} in the
general case. These methods extend to all of the other exceptional
Lie groups. We then go on to describe a one-parameter family of
$L$-functions over a function field, whose value distribution
coincides with that of the characteristic polynomials associated
with the 7-dimensional representation of $G_2$ in the limit as the
size of the finite field grows (this was proved by Katz
\cite{Katz-expsums}). The link with finite fields is natural,
because $N$ is fixed for the exceptional groups and the
$L$-functions in question (whose zeros correspond to eigenvalues)
are polynomials.

This note is organized as follows.  In Section 2 we review
properties of $G_2$ necessary for our random matrix calculations.
These calculations are performed in Section 3.  The $L$-functions
associated with $G_2$ are constructed in Section 4.  In Section 5
we conclude with a brief discussion of the generalization to the
other exceptional Lie groups.

\section{Preliminaries about $G_2$}

As pointed out in the Introduction, the exceptional Lie groups are
closely related to the classical matrix groups. One particularly
natural way of seeing this relationship is via their Lie algebras.
An important class of Lie algebras, because they form building
blocks of more general algebras, is the class of complex
semi-simple Lie algebras. This class allows a complete
categorization and is elegantly summarized in the possible Dynkin
diagrams which encapsulate the allowed root systems; the root
systems describe the structure constants of the Lie algebra
(standard references for this material include
\cite{Fulton,Gilmore, Cornwell}).  The result of the analysis is
that the structure of possible root systems is highly constrained.
Indeed the only possibilities fall into four infinite families,
$a_n$, $b_n$, $c_n$ and $d_n$ plus five exceptional cases, $g_2$,
$f_4$, $e_6$, $e_7$ and $e_8$.  Each complex Lie algebra has a
compact real form and this real form is the Lie algebra of a
compact group; the compact real forms of $a_n$,  $b_n$, $c_n$ and
$d_n$ are the Lie algebras of $SU(n+1)$, $SO(2n+1)$, $Sp(2n)$ and
$SO(2n)$ respectively.  The five exceptional cases are the Lie
algebras of the groups $G_2$, $F_4$, $E_6$, $E_7$ and $E_8$.  What
is significant for our purposes is the many of the constructions
which exist for the classical groups have analogues for the
exceptional groups.

Here, for concreteness, we will mainly focus on the smallest
exceptional case, namely $G_2$, for which the Lie algebra is the
compact real form of $g_2$. The group $G_2$ is the automorphism group
of the octonions, and has an embedding into $SO(7)$.

The group is 14-dimensional and has rank 2.  The six positive
roots may be taken to be
\begin{eqnarray}
\alpha_1 &=& \left(
\begin{array}{c}
 1 \\
  0
\end{array}
\right); \quad \alpha_2 = \left(
\begin{array}{c}
 - {3/ 2} \\
  {\sqrt{3}/ 2}
\end{array}
\right);
\nonumber\\
\alpha_3 &=& \alpha_1 + \alpha_2= \left(
\begin{array}{c}
  {-1/ 2} \\
  {\sqrt{3}/ 2}
\end{array}
\right); \quad \alpha_4 = 2\alpha_1 +  \alpha_2= \left(
\begin{array}{c}
  {1/ 2} \\
  {\sqrt{3}/ 2}
\end{array}
\right);
\nonumber\\
\alpha_5 &=& 3\alpha_1 +  \alpha_2= \left(
\begin{array}{c}
 3/2 \\
  \sqrt{3}/2
\end{array}
\right); \quad \alpha_6 = 3\alpha_1 + 2\alpha_2= \left(
\begin{array}{c}
  {0} \\
  {\sqrt{3}}
\end{array}
\right).
\end{eqnarray}
The complete set of roots is $R=\{\pm \alpha_i \},\ {i=1...6}$.
The set of short roots is
$R_S=\{\pm \alpha_1, \pm \alpha_3,\pm \alpha_4\}$;
the set of long roots is
$R_L=\{\pm \alpha_2,\pm \alpha_5, \pm \alpha_6\}$.
The Weyl group $W$ of $G_2$ is a dihedral group with $12$ elements.

Let $T$ be a maximal torus of $G_2$, which is isomorphic to a product
of two circles $S^1\times S^1$.  Every element of $G_2$ is conjugate
in $G_2$ to an element of $T$, which is unique up to conjugation by
the Weyl group.
%Let $H_1$ and $H_2$ be the generators of the Cartan subalgebra of
%$G_2$.  Every element of $G_2$ is conjugate to one of the form
%$e^{i(\theta_1 H_1+ \theta_2 H_2)}$.

Weyl's integration formula reads: If $d\mu_{\rm inv}(g)$ is the Haar
probability measure on  $G_2$, $dt$ is the Haar probability measure on
$T$ and $F$ is a continuous function on $G_2$, invariant under
conjugation, then
\begin{eqnarray}
\int_{G_2} F(g) d\mu_{\rm inv}(g) =
{1\over 12} \int_T F(t) |\Delta(t)|^2 dt
\end{eqnarray}
%\int_{\theta_1=-2\pi}^{2\pi}\int_{\theta_2=-2\pi/\sqrt{3}}^{2\pi/\sqrt{3}}{d\theta_1\over
%4\pi}{d\theta_2\sqrt{3}\over 4\pi} ( F| \Delta|^2 )(e^{i(\theta_1 H_1+ \theta_2 H_2)} )
%\end{eqnarray}
where
\begin{eqnarray}
\Delta(t) = \sum_{\sigma\in W} (\det \sigma)
t^{\sigma(\delta)} = t^{\delta} \prod_{\alpha>0}
(1-t^{-\alpha})
\end{eqnarray}
%\begin{eqnarray}
%\Delta(\theta_1,\theta_2) := \sum_{\sigma\in W} (\det \sigma)
%e^{i\sigma(\delta).\Theta} = e^{i\delta.\Theta} \prod_{\alpha>0}
%(1-e^{-i\alpha.\Theta})
%\end{eqnarray}
(the equality is Weyl's denominator formula),
where $\delta$ is half the sum of the positive roots:
\begin{eqnarray}
\delta={1\over 2}\sum_{\alpha>0} \alpha = 5\alpha_1 + 3 \alpha_2.
\end{eqnarray}
and  $\sigma(\delta)$ means the vector obtained from $\delta$ by
the Weyl group element $\sigma$. If we parametrize a particular
maximal torus by $t=(t_1,t_2)$, then $t^\alpha$ is an expression
of the form $t_1^{\alpha.e_1}t_2^{\alpha.e_2}$ where $e_1$ and
$e_2$ are (two-component) vectors; the vectors $e_1$ and $e_2$ and
range of the parameters $t_1$ and $t_2$   depend on the choice of
maximal torus. We derive our results below without needing to make
an explicit choice for this torus.

For any pair of integers $[n_1,n_2]$ there is an irreducible
representation $\rho_{[n_1,n_2]}$ which has highest weight
$\lambda_{[n_1,n_2]} = n_1 \omega_1 + n_2 \omega_2$ where
$\omega_1=\alpha_4$ and $\omega_2=\alpha_6$ are the fundamental
weights.

The character $\chi_\lambda$ of the representation
evaluated at the group element $t\in T$ is
%$e^{i(\theta_1 H_1+ \theta_2 H_2)}$
\begin{eqnarray}
\chi_\lambda(t) = {\rm Tr} \left[ \rho_\lambda(t)\right]
= \sum_\mu d_\mu t^{\mu},
\end{eqnarray}
where $\mu$ are the  weights of the representation, $d_\mu$ is the
multiplicity of the weight $\mu$.

The character $\chi_\lambda(t)$ and dimension
$d_\lambda$ of the representation $\rho_\lambda$ are given by
Weyl's formulae:
\begin{eqnarray}
\chi_\lambda(t)&=&{\sum_{\sigma\in W} (\det
\sigma) t^{\sigma(\lambda + \delta)} \over \sum_{\sigma\in
W} (\det \sigma) t^{\sigma( \delta)} }; \label{Weyl}\\
d_\lambda&=&{\prod_{\alpha> 0} (\lambda + \delta).\alpha \over
\prod_{\alpha> 0} \delta.\alpha}.
\end{eqnarray}
The sums in (\ref{Weyl}) are over elements $\sigma$ in the Weyl
group.
With these conventions, the orthogonality relation for characters
is
\begin{eqnarray}
{1\over 12}
\int_T
%\int_{\theta_1=-2\pi}^{2\pi}\int_{\theta_2=-2\pi/\sqrt{3}}^{2\pi/\sqrt{3}}{d\theta_1\over
%%4\pi}{d\theta_2\sqrt{3}\over 4\pi}
| \Delta(t)|^2\ \chi_{[n_1,n_2]}(t)
\overline{\chi_{[m_1,m_2]}(t)}\ dt =
\delta_{n_1,m_1}\delta_{n_2,m_2}.
\end{eqnarray}

We will be particularly interested in the fundamental representations
$[1,0]$   (induced from the embedding of $G_2$ into $SO(7)$),
and $[0,1]$ (the adjoint representation),
which have characters
\begin{eqnarray}
\chi_{[1,0]}(t) &=&
%e^{i(\theta_1+\sqrt{3}\theta_2)/2} +
%e^{i(-\theta_1+\sqrt{3}\theta_2)/2}\nonumber \\
%& &\quad+ e^{i(-\theta_1-\sqrt{3}\theta_2)/2} +
%e^{i(\theta_1-\sqrt{3}\theta_2)/2}
%\nonumber\\
%& &\quad + e^{i\theta_1} + e^{-i\theta_1}
%+1 \nonumber\\
 1+  \sum_{\alpha\in R_S} t^{\alpha};  \label{character[1,0]} \\
\chi_{[0,1]}(t) &=&
%e^{i(\theta_1+\sqrt{3}\theta_2)/2} +
%e^{i(-\theta_1+\sqrt{3}\theta_2)/2}
%\nonumber\\
%& &\quad+ e^{i(-\theta_1-\sqrt{3}\theta_2)/2} +
%e^{i(\theta_1-\sqrt{3}\theta_2)/2}
%\nonumber\\
%& &\quad + e^{i(3\theta_1+3\sqrt{3}\theta_2)/2} +
%e^{i(-3\theta_1+3\sqrt{3}\theta_2)/2}
%\nonumber\\
%& &\quad + e^{i(-3\theta_1-3\sqrt{3}\theta_2)/2} +
%e^{i(3\theta_1-3\sqrt{3}\theta_2)/2}
%\nonumber\\
%& &\quad + e^{i\theta_1} + e^{-i\theta_1} + e^{i\sqrt{3}\theta_2}
%+ e^{-i\sqrt{3}\theta_2}+2
%\nonumber\\
  2+  \sum_{\alpha\in R} t^{\alpha};
\label{character[0,1]}
\end{eqnarray}
and dimensions
\begin{eqnarray}
d[1,0]=7;\quad d[0,1]=14.
\end{eqnarray}

\section{Characteristic polynomials}

We will focus on the characteristic polynomials
\begin{eqnarray}
Z(U_\rho,\theta):=\det(1-U_\rho e^{-i\theta})
\end{eqnarray}
of (unitary) matrices $U_\rho$ coming from a given representation
$\rho$ of the group.  The group elements which these matrices
represent can be thought of as being chosen randomly from the
group with respect to the uniform invariant (Haar) measure.  We
will calculate explicit expressions for the averages (over the
group, with respect to Haar measure) of $|Z|^s$ for complex
numbers $s$ (see \cite{Keating-Snaith1, Keating-Snaith2} for
analogous calculation relating to $U(N)$, $O(N)$, and $USp(2N)$).

First let us consider the modulus of $Z$.  We wish to calculate
\begin{eqnarray}
<|Z(U_\rho,\theta)|^s>_{G_2} &=& \int |Z(U_\rho(g),\theta)|^s\nonumber d\mu_{\rm
inv}(g)\\
&=& \int |\det(1-U_\rho(g) e^{-i\theta})|^s d\mu_{\rm
inv}(g).
\end{eqnarray}
Since the integrand is a class function, this integral reduces to
an integral over the  maximal torus $T$:
\begin{eqnarray}
 <|Z(U_\rho,\theta)|^s>_{G_2} =
 {1\over 12} \int_T
| \Delta(t)|^2
|\det(1-U_\rho(t)
e^{-i\theta})|^s dt .   \label{integral1}
%\int_{\theta_1=-2\pi}^{2\pi}\int_{\theta_2=-2\pi/\sqrt{3}}^{2\pi/\sqrt{3}}{d\theta_1\over4\pi}{d\theta_2\sqrt{3}\over
%4\pi}\
\end{eqnarray}

\subsection{The seven-dimensional representation}

This representation is induced by the embedding of $G_2$ as a subgroup
 of $SO(7)$.
 From (\ref{character[1,0]}), we can calculate that
\begin{eqnarray}
 Z(U_{[1,0]},\theta)=\det(1-U_{[1,0]}(t) e^{-i\theta} ) =
 (1-e^{-i\theta})\prod_{\alpha\in R_S} (1- t^{\alpha}e^{-i\theta}).
\end{eqnarray}
We see that $Z$ has a zero at $\theta=0$,
as is the case for the characteristic polynomial of any
odd-dimensional orthogonal matrix.
Let us define $\hat Z$ as
\begin{eqnarray}
 \hat Z(U_{[1,0]},\theta)=
 (1-e^{-i\theta})^{-1} Z(U_{[1,0]},\theta).
 \end{eqnarray}
   We now present a
 formula for $<| \hat Z(U_{[1,0]},\theta) |^s>$ at $\theta=0$.

We have that
\begin{eqnarray}
 <|\hat Z(U_{[1,0]},0)|^s>_{G_2}
& =& {1\over 12 }\int_T  |\Delta(t)|^2
|\prod_{\alpha\in R_S}(1-t^{\alpha}) |^s dt \nonumber \\
 &= & {1\over 12 }\int_T
|\Delta(t)|^2 \ \prod_{\alpha\in R_S}(1- t^{\alpha} )^s dt
,\label{integral2}
\end{eqnarray}
but
\begin{eqnarray}
 |\Delta(t)|^2 = \prod_{\alpha\in R}(1- t^{\alpha}),
\end{eqnarray}
and so
\begin{eqnarray}
<|\hat Z(U_{[1,0]},0)|^s>_{G_2} =  {1\over 12 } \int_T
\prod_{\alpha\in R}(1- t^{\alpha})^{k_\alpha}dt ,
\label{integral3}\label{int1}
\end{eqnarray}
where
\begin{eqnarray}
k_\alpha
 = \cases{
    s+1 & {\rm if $\alpha\in R_S$},\cr
    1 & {\rm if $\alpha\in R_L$} \cr}.
\end{eqnarray}

Consider first the case that $s$ is an integer.  The value of the
integral (\ref{integral3}) is  then the constant term in the
expression
\begin{eqnarray}
{1\over 12}\prod_{\alpha\in R}(1- t^{\alpha})^{k_\alpha}.
\end{eqnarray}
The value of this constant term is in turn one of Macdonald's
celebrated constant term conjectures \cite{Macdonald}, proved
for $G_2$ by Zeilberger \cite{Zeilberger} and Habsieger \cite{Habsieger} (see
Opdam \cite{Opdam} for a uniform proof):
\begin{eqnarray}
{ (3k_S + 3 k_L)!(2k_S)!(2 k_L)!( 3 k_L)! \over 12 (2k_S + 3
k_L)!( k_S+ 2k_L)!(k_S+ k_L)! (  k_S)! ((  k_L)!)^2 },
\end{eqnarray}
where $k_S$ (resp. $k_L$) is the value of $k_\alpha$ for the short
(resp. long) roots.

Thus for the representation $[1,0]$, $k_S = s+1$ and $k_L =1$, and
so for $s$ a positive integer or zero,
\begin{eqnarray}
<|\hat Z(U_{[1,0]},0)|^s>_{G_2} = { (3s+6)!(2s+2)! \over
 (2s+ 5)!( s+3)!(s+2)! (  s+1)!
}.
\end{eqnarray}

It follows from Carlson's theorem (see \cite{Titchmarsh}) that
then
\begin{eqnarray}
<|\hat Z(U_{[1,0]},0)|^s>_{G_2} = { {\rm \Gamma}(3s+7){\rm
\Gamma}(2s+3) \over
 {\rm \Gamma}(2s+ 6){\rm \Gamma}( s+4){\rm \Gamma}(s+3){\rm \Gamma}(  s+2)
}\label{gamma1}
\end{eqnarray}
for ${\rm Re}s>-3/2$.  To see this, note that one may deduce
directly from (\ref{int1}) that $2^{-6s}<|\hat
Z(U_{[1,0]},0)|^s>_{G_2}$ is bounded when ${\rm Re}s>0$, and from
Stirling's formula that
\begin{equation}
2^{-6s}{ {\rm \Gamma}(3s+7){\rm \Gamma}(2s+3) \over
 {\rm \Gamma}(2s+ 6){\rm \Gamma}( s+4){\rm \Gamma}(s+3){\rm \Gamma}(  s+2)
}
\end{equation}
is also bounded in the same half-plane.  The function
\begin{equation}
2^{-6s}\left(<|\hat Z(U_{[1,0]},0)|^s>_{G_2}-{ {\rm
\Gamma}(3s+7){\rm \Gamma}(2s+3) \over
 {\rm \Gamma}(2s+ 6){\rm \Gamma}( s+4){\rm \Gamma}(s+3){\rm \Gamma}(  s+2)
}\right)
\end{equation}
is therefore regular and bounded in ${\rm Re}s>0$, and vanishes
when $s$ is a non-negative integer.  Carlson's theorem therefore
implies (\ref{gamma1}).

Now that (\ref{gamma1}) has been established, we may immediately
write down expressions for the probability density functions
associated with the value distributions of $\log|\hat
Z(U_{[1,0]},0)|$,
\begin{eqnarray}
P_1(x)=\frac{1}{2\pi}\int_{-\infty}^{\infty}{ {\rm
\Gamma}(3iy+7){\rm \Gamma}(2iy+3) \over
 {\rm \Gamma}(2iy+ 6){\rm \Gamma}( iy+4){\rm \Gamma}(iy+3){\rm \Gamma}(  iy+2)}{\rm e}^{-iyx}dy
\end{eqnarray}
and $|\hat Z(U_{[1,0]},0)|$,\label{P1}
\begin{eqnarray}
P_2(x)=\frac{1}{2\pi ix}\int_{c-i\infty}^{c+i\infty}{ {\rm
\Gamma}(3s+7){\rm \Gamma}(2s+3) \over
 {\rm \Gamma}(2s+ 6){\rm \Gamma}( s+4){\rm \Gamma}(s+3){\rm \Gamma}(  s+2)
}x^{-s}ds\label{P2}
\end{eqnarray}
for any $c>0$.  One can easily deduce asymptotic properties of the
probability density functions from these integrals; see, for
example \cite{Keating-Snaith1, Keating-Snaith2}.

We note finally that $\hat Z$ is real and positive at $\theta =0$
and so its phase there is zero.

\subsection{The fourteen-dimensional representation}

In this case the determinant $Z$ has a double zero at $\theta=0$
(corresponding to the twice repeated weight 0, see
(\ref{character[0,1]})).  Thus we define
\begin{eqnarray}
\hat Z(U_{[0,1]},\theta) = (1-e^{-i\theta})^{-2} \det(1-U_{[0,1]}
e^{-i\theta}).
\end{eqnarray}
Similar calculations to the previous case show that for $s$ a
positive integer or zero, $<|\hat Z(U_{[0,1]},0)|^s >_{G_2}$ is
given by the constant term in
\begin{eqnarray}
{1\over 12}\prod_{\alpha\in R}(1- t^{\alpha})^{k_\alpha},
\end{eqnarray}
where in this case,
\begin{eqnarray}
k_S = k_L= s+1.
\end{eqnarray}
Thus for $s$ a non-negative integer the Macdonald identity quoted
above implies that
\begin{eqnarray}
<|\hat Z(U_{[0,1]},0)|^s>_{G_2} = { (6s+6)!(2s+2)! \over 12 (5s+
5)! ((  s+1)!)^3 }.
\end{eqnarray}

Once again, Carlson's theorem may be applied (in this case after
multiplication by $2^{-12s}$) to show that
\begin{eqnarray}
<|\hat Z(U_{[0,1]},0)|^s>_{G_2} = { {\rm \Gamma} (6s+7){\rm
\Gamma}(2s+3) \over 12 {\rm \Gamma}(5s+ 6) \left[{\rm \Gamma}(
s+2)\right]^3 }
\end{eqnarray}
for ${\rm Re}s>-7/6$.  This can then be used to write down
expressions for the probability density functions associated with
the value distributions of $\log|\hat Z(U_{[0,1]},0)|$ and $|\hat
Z(U_{[0,1]},0)|$, as in the previous section.

Also, as in the case of the representation $[1,0]$, the phase of
$\hat Z$ is zero.

\subsection{Other representations and $\theta \neq 0$}

For other representations of $G_2$ and for $\theta \neq 0$, the
integrand is not of the form
\begin{eqnarray}
\prod_{\alpha\in R}(1- t^{\alpha})^{k_\alpha},
\end{eqnarray}
%$k_\alpha=k_\beta$ if $|\alpha|=|\beta|$
and, as far as we are aware, closed form expressions for these
integrals are not known.

\section{Value distribution problems over function fields}

Our purpose now is to outline the analogy between number fields
and function fields over a finite field in  the context of the
value distribution of zeta- and $L$-functions for these cases and
the predicted behaviour in terms of Random Matrix Theory.

It was proved by Selberg that the logarithm of the Riemann zeta
function on the critical line has a Gaussian value distribution
\cite{Selberg-zeta},  and the same is true for all $L$-functions
\cite{Amalfi} under suitable assumptions. Selberg also
investigated the ``$q$-analogue'' of these results for the value
distribution of the family of Dirichlet $L$-functions
\cite{Selberg-dirichlet} at a point on the critical line and there
too obtained a Gaussian value distribution. Precisely, for $q$
prime we have $q-2$ primitive characters $\chi$ modulo $q$, and
for fixed $t$ consider the $q-2$  numbers
$$
\frac {arg L(\frac 12 +it,\chi)}{\sqrt{\frac 12 \log\log q}}
$$
($\chi$ varies over all primitive/nonprincipal characters modulo
$q$). Then as $q\to \infty$, these numbers are distributed as a
standard Gaussian.

Our purpose in this section is two-fold: first it is to point out
that there are  corresponding results for various families of
$L$-functions over function fields, with the r\^{o}le of the
Gaussian being replaced by various distributions from Random
Matrix Theory; and second to construct an example for which the
appropriate random-matrix distribution is the $G_2$ result
calculated above.

Likewise, there are similar results for the {\em moments} of the
$L$-functions, which in the number field setting are mostly
conjectural \cite{Keating-Snaith1, Keating-Snaith2, CFKRS}, but in
the function field setting can sometimes be proved.

\subsection{Zeta functions}

Let $k$ be a finite field of cardinality $q$ and  $X/k$ a (smooth,
geometrically connected, proper) curve defined over $k$. The zeta
function of $X/k$ is given by the series
$$
Z(X;T) = \exp(\sum_{n=1}^\infty N_n \frac{T^n}n),
$$
where $N_n = \#X(k_n)$ is the number of points of $X$ over the
field $k_n$, the extension of $k$ of degree $n$.  The series is
absolutely convergent for $|T|<1$.

Trivial example: take  $X=\mathbf P^1$, the projective line. The
number of points of $\mathbf P^1$ over the finite field $k_n$ is
$\#k_n+1=q^n+1$ and so
$$
Z(\mathbf P^1; T)  = \frac 1{(1-qT)(1-T)}.
$$

This zeta function has an Euler product
$$
Z(X;T) = \prod_p (1-T^{\deg p})^{-1} ,\qquad |T|<1,
$$
where $p$ runs over all closed points of $X$. (In the example of
$\mathbf P^1$, the closed points $p$ correspond to irreducible
monic polynomials  $p(x)\in k[x]$ with the addition of the ``point
at infinity''.)

It turns out that $Z(X;T)$ is a {\em rational} function of $T$, of
the form
$$Z(X;T) = \frac{P(X;T)}{(1-T)(1-qT)} $$
with $P(X;T)\in 1+T\mathbf Z[T]$ a monic integer polynomial of
degree $2g$, $g$ being the genus of the curve $X$, which we can
write as $P(X;T) = \prod_{j=1}^{2g}(1-\alpha_j T)$. The inverse
roots $\alpha_j$ are thus algebraic integers. Further, there is a
functional equation $T\mapsto q/T$:
$$
Z(X;\frac 1{qT}) =  q^{1-g}T^{2-2g} Z(X;T).
$$
If we set $T=q^{-s}$ then the functional equation translates into
$s\mapsto 1-s$.

The ``Riemann Hypothesis for curves over a finite field'' (proved
in the general case by A. Weil) is that all the inverse roots
$\alpha_j$ of $P(X;T)$ have absolute value $\sqrt{q}$, that is as
a function of the variable $s$ all zeros are on the line ${\rm
Re}s=1/2$.

What is especially important for our purpose is that the
polynomial $P(X;T)$ is the characteristic polynomial of a matrix:
%$\det(I-T F_X| V)$ of the ``Frobenius
%automorphism of $X$'' acting on a certain vector space $V$. This
%vector space is the cohomology group  $V=H^1(X\otimes \bar k,\mathbf
%{\bar Q_\ell})$
%where $\mathbf {\bar Q_\ell}$ is the algebraic closure of the field
%of $\ell$-adic numbers, with $\ell$ a prime distinct from the
%characteristic of $k$.
there is a unique conjugacy class $\Theta_X\in USp(2g)$ in the
unitary symplectic group   such that $P(X;T) =
\det(I-q^{1/2}T\Theta_X)$.

\subsection{Families of curves}

Now consider a ``family'' of curves $X/k$. In order to study the
behaviour of $P(X;T)$ as $X$ varies, it suffices to understand the
distribution of the conjugacy classes $\Theta_X$. In several
cases, it is known that as $q\to \infty$ these become
equidistributed in $USp(2g)$ (with respect to Haar measure).

For instance, this is the case for the family $\mathcal M_g$ of
{\em all} $k$-isomorphism classes of (smooth, geometrically
connected, proper) curves of given genus $g$ \cite[Theorem
10.7.15]{KS}.

This allows one to compute arithmetic quantities such as the {\em
moments} of $P(X;T)$ as $X$ varies in $\mathcal M_g(k)$ by using
the corresponding (non-arithmetic) computation in Random Matrix
Theory for $USp(2g)$. Thus one finds that for $T$ fixed, say
$q^{-1}T=1$, one has
$$
\lim_{q\to \infty} \frac 1{\#\mathcal M_g(k)} \sum_{X\in \mathcal
M_g(k)} P(X,q^{1/2})^s = \int_{USp(2g)} \det(I-A)^s d_{Haar}(A).
$$

The moments of the characteristic polynomial in $USp(2g)$ were
computed in \cite{Keating-Snaith2} and are given by
$$
\int_{USp(2g)} \det(I-A)^s d_{Haar}(A) =2^{2gs} \prod_{j=1}^g
\frac{\Gamma(1+g+j)\Gamma(1/2+s+j)}{
\Gamma(1/2+j)\Gamma(1+s+g+j)}.
$$
The probability density functions for the value distributions
associated with the polynomial and its logarithm may then be
written as integrals, as in (\ref{P1}) and (\ref{P2})
\cite{Keating-Snaith2}.  In the case of the logarithm of the
characteristic polynomial, the limit distribution when
$g\rightarrow \infty$ is a Gaussian.

\subsection{$L$-functions attached to exponential sums}

We consider one-variable exponential sums constructed as follows:
let $k$ be a finite field with $q$ elements as above, $f(x)$ and
$h(x)\in k(x)$ be rational functions, $\psi$ a nontrivial additive
character of $k$ (e.g. for $k=\Z/ p\Z$ take $\psi(x)=\exp(2\pi i
ax/p)$, $0\neq a \in \Z/p\Z$), and $\chi$ a  multiplicative
character of $k^\times$. Set
$$
S(\psi,\chi;f,h;q) = \sum_x \psi(f(x))\chi(h(x)),
$$
the sum running over all $x\in k$ which are not poles of $f,h$ and
such that $h(x)\neq 0$. For the finite extension $k_n$ of degree
$n$ of $k$, we get nontrivial characters $\psi_n= \psi\circ
\Tr_{k_n/k}$ and $\chi_n = \psi\circ \Norm_{k_n/k}$ by composing
with the trace and norm maps. Correspondingly we get exponential
sums for $k_n$
$$
S_n(\chi,\psi;f,h):=S(\chi_n,\psi_n;f,h;q^n).
$$
The $L$-function is defined as
$$
L(S,T) = \exp(\sum_{n=1}^\infty S_n(\chi,\psi;f,h) \frac{T^n}n ).
$$
These have an Euler product decomposition and are rational
functions of $T$.

In many cases of interest to us, it turns out that $L(T)$ is in
fact a polynomial  of the form $\det(I-q^{1/2}\Theta_S)$ with
$\Theta_S$ a unitary matrix.

\subsection{ Gauss sums}
Given a nontrivial additive character $\psi$ of $k$ and a
nontrivial multiplicative character $\chi$ of $k^\times$, one
defines the Gauss sum $g(\chi,\psi)$ by
$$
g(\chi,\psi) = \sum_{x\neq 0} \chi(x)\psi(x).
$$
%For the finite extension $k_n$ of degree $n$ of $k$, we get nontrivial
%characters $\psi_n= \psi\circ \Tr_{k_n/k}$ and $\chi_n = \psi\circ
%\Norm_{k_n/k}$ by composing with the trace and norm maps.
Correspondingly we get Gauss sums for $k_n$
$$
g_n(\chi,\psi):=g(\chi_n,\psi_n).
$$
To compute the $L$-function
$$
L(g(\chi,\psi),T) = \exp(\sum_{n=1}^\infty g_n(\chi,\psi)
\frac{T^n}n)
$$
one can use the Hasse-Davenport relations:
$$
-g_n(\chi,\psi) = (-g(\chi,\psi))^n.
$$
These give
$$
L(g(\chi,\psi),T) = 1+Tg(\chi,\psi).
$$

As is well known, $|g(\chi,\psi)|=\sqrt{q}$. Thus we may write
$g(\chi, \psi) = \sqrt{q} e^{i\theta_\chi}$ (we omit the
dependence on $\psi$ which is of a trivial nature). Setting
$s=1/2+i\theta/\log q$, $T=q^{-s} = q^{-1/2}e^{-i\theta}$, we find
$$ L(g(\chi,\psi),T) = 1+e^{i(\theta_\chi-\theta)}.$$

The $q-2$ angles $\{\theta_\chi:\chi\neq \chi_0 \}$ are uniformly
distributed in $[0,2\pi)$. This is easy to see from Deligne's
estimate on hyper-Kloosterman sums, see \cite[section
1.3.3]{Katz-asterisque}. Thus the moments of $L(g(\chi,\psi),T)$
and of its logarithm, averaged over $\chi$ and taken as $q\to
\infty$, are the same as those for the function  $1+e^{i\theta}$.

\subsection{Kloosterman sums (i)}

These are the sums
$$
\kl(a,p) = \sum_{x_1 x_2=a \bmod p}\exp \frac {2\pi
i}{p}(x_1+x_2).
$$
More generally for a finite field $k$ with $q$ elements, take a
nontrivial additive character $\psi$ and $a\neq 0$, and set
$$
\kl(a,q) = \sum_{x_1 x_2=a} \psi(x_1+x_2).
$$
This sum is real (replace $x\mapsto -x$), and as Weil proved
satisfies
$$
|\kl(a,q)|\leq 2\sqrt{q}\;.
$$

The associated $L$-function is a polynomial of degree $2$
$$
L(\kl(a,q),T) = 1 + \kl(a,q) T +q T^2.
$$
It is of the form $\det(I-q^{1/2}\Theta_a)$ with $\Theta_a\in
SU(2)$. It was shown by Katz \cite{Katz-gauss} that as
$q\to\infty$, the $q-1$ conjugacy classes $\{\Theta_a: a\in
k^\times\}$ become equidistributed in $SU(2)$ with respect to Haar
measure. This implies we can compute the value distribution of $L$
and $\log L$ via RMT on $SU(2)$.

\subsection{Kloosterman sums (ii)}

We next look at an example of exponential sums in several
variables: hyper-Kloosterman sums are $n$-variable sums ($n\geq
2$) generalizing the previous example, given by
$$
\kl_n(a,q) = \sum_{x_1 x_2\dots x_n=a} \psi(x_1+x_2+\dots +x_n).
$$
Replacing $\psi$ by $\psi\circ \Tr_{k_m/k}$ gives the sum
$\kl_n(a,q^m)$.

The associated $L$-function is defined as
$$
L_a(T):=\exp \left((-1)^{n} \sum_{m=1}^\infty \kl_n(a,q^m)
\frac{T^m}m \right).
$$
This $L$-function is a polynomial of degree $n$. It was proved by
Katz \cite{Katz-gauss} that  it  can be written as $\det(I-q^{(n-1)/2} T
\Theta_n(a,q))$, with $\Theta_n(a,q)\in K_n$, where $K_n$ is the
compact group $USp(2n)$, $n$  even, $SU(n)$, $n$, $q$  odd,
$SO(n)$, $q$ even, $n\neq 7$ odd, and $G_2$,  $q$ even , $n=7$.
Moreover, as $a$ varies through all nonzero elements of $k$, the
$q-1$ conjugacy classes $\Theta_n(a,q)$ of $K_n$ become
equidistributed there as $q\to \infty$ while keeping the type of
$K_n$ fixed. For instance, taking $n=7$ and $q=2^r$, $r\to \infty$
gives $2^r-1$ conjugacy classes $ \{ \Theta_7(a,2^r):0\neq a\in
\mathbf F_{2^r}\} $  which become equidistributed in $G_2$ as
$r\to \infty$.

\subsection{An exponential sum associated to $G_2$}

Let $p$ be a prime, $p\geq 17$, $k=\Z/p\Z$, $\chi_{(2)}$ the
unique quadratic character (Legendre symbol) of $k^\times$, and
$\psi$ a nontrivial additive character of $k$, that is $\psi(x) =
e^{2\pi i ax/p}$ for some $a\in k^\times$. Consider for $t\in
k^\times$ the exponential sum
$$
\nmk(t) = \sum_{x\in  k^\times} \chi_{(2)}(x) \psi(x^7+tx).
$$
These sums were studied by Nick Katz  and the results below are
due to him \cite{Katz-expsums}.

Note that $\overline{\nmk(t)} = \chi_{(2)}(-1)\nmk(t)$ and so
$\nmk(t)$ is real if $\chi_{(2)}(-1)=1$, that is if $p=1 \bmod 4$,
and imaginary if $\chi_{(2)}(-1)=-1$, i.e. if $p=3 \bmod 4$.

%$\nmk(\zeta t) = \nmk(t)$ for $\zeta$ any $7$-th root of unity. These
%exist (other than $1$) iff $p=1 \bmod 7$.

In view of the transformation properties under complex
conjugation, we divide  the exponential sum $\nmk(t)$ by the
quadratic Gauss sum $g(\chi_{(2)})$ to get a {\em real} number.
Furthermore, there is a (unique) choice of sign $\epsilon_p=\pm 1$
so that\footnote{At the time of writing we do not know how to
determine $\epsilon_p$.}
$$
\nmk'(t) = \epsilon_p \frac{\nmk(t)}{g(\chi_{(2)})}
$$
is minus the trace of a matrix  $\Theta_t \in SO(7)$: $\nmk(t) =
-\Tr\Theta_t$. Moreover, this matrix turns out to lie in $G_2$.

The associated $L$-function is a polynomial of degree $7$, which
is a characteristic polynomial of the element $\Theta_t$ of $G_2$,
$$
L(\nmk'(t),T) = \det(I-\Theta_t T).
$$
As $t$ varies in $(\Z/p/Z)^\times$, these $p-1$ conjugacy classes
$\Theta_t$ become equidistributed in $G_2$ as $p\to \infty$. Thus
the value distribution of $L(\nmk'(t),T)$ at fixed $T$ is computed
by RMT for $G_2$.

\section{Other Lie groups}

The random matrix calculations reported here for $G_2$ generalize
straightforwardly to the other exceptional Lie groups.  In each
case one has a Weyl integration formula, which allows the moments
of the characteristic polynomials associated with representations
of the group to be written as integrals over the Cartan subgroup,
and a Macdonald identity, which enables the integrals to be
evaluated as ratios of ${\rm \Gamma}$-functions.  This prompts the
question as to whether families of finite-field $L$-functions can
be constructed whose value distributions are given by each of the
other exceptional Lie groups.

As a final remark, we note that in \cite{Macdonald} Macdonald
gives constant term formulae for affine root systems (which are
related to Kac-Moody algebras).  This suggests the intriguing
possibility of extending the ideas described in this paper to
families of random matrices arising from the representations of
the associated infinite dimensional groups.

\section*{Acknowledgments}

We are grateful to Nick Katz for a very helpful correspondence.
JPK is also grateful to the American Institute of Mathematics for
generous hospitality during the completion of this manuscript.

\end{document}